\documentclass[aps,pre,10pt,groupedaddress,amsfonts,amssymb,amsmath,showpacs,showkeys,a4paper,floatfix,notitlepage]{revtex4-1}

\usepackage[english]{babel}
\usepackage[T1]{fontenc}
\usepackage{lmodern}
\usepackage[utf8]{inputenc}
\usepackage[protrusion=true,expansion=true]{microtype}
\usepackage{amsmath,amsfonts,amsthm}
\usepackage{ bbold }
\usepackage[pdftex]{graphicx}
\usepackage[outdir=./]{epstopdf}
\usepackage{bm}
\usepackage[usenames,dvipsnames,svgnames,table]{xcolor}
\usepackage{csquotes}
\usepackage{booktabs} 
\usepackage{array} 
\usepackage{geometry}
\usepackage{MnSymbol}
\usepackage{pgf}
\usepackage{import}
\usepackage{caption}
\usepackage{subfigure}
\usepackage{wrapfig} 
\usepackage[normalem]{ulem} 
\usepackage{comment} 

\newif\ifchanges
\changestrue

\definecolor{lbgk_color}{rgb}{0,0,0}
\definecolor{kbcA_color}{HTML}{377EB8}
\definecolor{kbcB_color}{HTML}{4DAF4A}
\definecolor{kbcC_color}{HTML}{984EA3}
\definecolor{kbcD_color}{HTML}{FF7F00}
\definecolor{entropic_color}{HTML}{E41A1C}

\begin{document}

\title{Drops bouncing off macro-textured superhydrophobic surfaces}


\author{Ali Mazloomi Moqaddam}
\affiliation{Department of Mechanical and Process Engineering, ETH Zurich, 8092 Zurich, Switzerland}

\author{Shyam S. Chikatamarla}
\affiliation{Department of Mechanical and Process Engineering, ETH Zurich, 8092 Zurich, Switzerland}

\author{Ilya Karlin}
\email{karlin@lav.mavt.ethz.ch}
\affiliation{Department of Mechanical and Process Engineering, ETH Zurich, 8092 Zurich, Switzerland}

\date{\today}

\begin{abstract} 	
Recent experiments with droplets impacting a macro-textured superhydrophobic surfaces revealed new regimes of bouncing with a remarkable reduction of the contact time. 
We present here a comprehensive numerical study that reveals the physics behind these new bouncing regimes and quantify the role played by various external and internal forces that effect the dynamics of a drop impacting a complex surface.
For the first time, three-dimensional simulations involving macro-textured surfaces are performed. Aside from demonstrating that simulations reproduce experiments in a quantitative manner, the study is focused on analyzing the flow situations beyond current experiments. We show that the experimentally observed reduction of contact time extends to higher Weber numbers, and analyze the role played by the texture density.
Moreover, we report a non-linear behavior of the contact time with the increase of the Weber number for application relevant imperfectly coated textures, and also study the impact on tilted surfaces in a wide range of Weber numbers.
Finally, we present novel energy analysis techniques that elaborate and quantify the interplay between the kinetic and surface energy, and the role played by the dissipation for various Weber numbers.\end{abstract}

\maketitle

\section{Introduction}
\label{sec:intro}

Impact of liquid drops on solid surfaces is a beautiful and fascinating fluidics problem in physics, whose complexity derives from the possible co-existence of a variety of phenomena, occurring at multiple temporal and spatial event scales \cite{Roisman2009,Roisman2009a,yarin2006drop}. These include but are not limited to splash \cite{Mani2010,Mandre2011,Riboux2014,Xui2005,Xui2007}, 
phase-change induced surface levitation \cite{Antonini2013,Wachters1966,Biance2003,Tran2012,Tran2013}, 
skating \cite{Kolinski2012}, 
rebounding \cite{deRuiter2014,richard2002surface,bird2013reducing,liu2014pancake,Antonini2013a}, 
prompt tumbling-rebound \cite{Antonini2016} and trampoline effect \cite{Schutzius2015Nature}.
Surfaces with special wetting properties have profound implications in engineering including power generation, transportation, water desalination, oil and gas production, and microelectronics thermal management. 
Particularly interesting are surfaces with extreme wetting properties, which are efficient at either repelling (hydrophobic) or attracting liquids (hydrophilic) such as water and oils but can also prevent formation of biofilms or ice \cite{rykaczewski2012direct}. 
The degree of surface wetting, typically measured by a drop's equilibrium contact angle, depends on the balance of the products of corresponding interfacial surface areas and surface energies. 
From a theoretical perspective, the contact angle of a liquid interacting with a flat solid is predicted by Young's equation. 
However, in order to achieve extreme wetting properties, the interface between the droplet and the substrate must be structured and often contains an additional gas or liquid phase \cite{rykaczewski2012direct}. For example, nano- and/or microscale roughening of a flat hydrophobic substrate yields a super-hydrophobic surface (contact angle above $\theta=150^\circ$ and negligible contact angle hysteresis) through trapping gas. 
Drop repellence from hydrophobic and super-hydrophobic surfaces is an area of active research 
\cite{yarin2006drop,rein1993phenomena,jung2012mechanism,blossey2003self,tuteja2007designing,okumura2003water,
richard2002surface,bird2013reducing,liu2014pancake,Liu2015NatCom,Gauthier2015,Schutzius2015Nature}.

The time during which the drop stays in a contact with the solid after impact is termed contact time (or rebound time). 
Minimization of contact time is a central point in a rational design of hierarchically structured surfaces and has been the focus of recent studies 
\cite{Bird2002,Schutzius2014,richard2002surface,bird2013reducing,liu2014pancake,Liu2015NatCom,Gauthier2015,Schutzius2015Nature}.
Richard, Clanet, and Qu\'{e}r\'{e} \cite{richard2002surface} found that the conventional mechanism of rebound on macroscopically flat superhydrophobic surfaces (impact-spread-recoil-rebound, \cite{richard2002surface,richard2000bouncing,wang2007impact,okumura2003water}) scales universally with the inertia-capillarity time, 
\begin{equation}\label{eq:tau}
\tau=\sqrt{\frac{\rho_l R_0^3}{\sigma}},
\end{equation} 
with $\rho_l$, $R_0$, and $\sigma$ the liquid density, drop radius, and surface tension, respectively.
This scaling,  $t_{\rm {contact}}/\tau\approx2.2$, is notably independent on the drop kinetic energy, and holds in a range of Weber numbers \cite{richard2002surface}, 
\begin{equation}\label{eq:We}
{\rm {We}}=\frac{\rho_l R_0 U_0^2}{\sigma},
\end{equation}
where $U_0$ is the impact velocity.
However, Bird et al \cite{bird2013reducing} demonstrated that by adding a macro-texture (few hundred micrometers) as a ridge on the flat surface, the contact time reduces by about $37\%$. 
Recently, Liu et al \cite{liu2014pancake} demonstrated that impact on a flat surface decorated with a lattice of sufficiently tall (almost a millimeter) tapered posts with a nanoscale superhydrophobic coating results in an unconventional mechanism where the drop rebounds even before the retraction takes place. Because of the flattened droplet shape at the time of rebound, this phenomenon was referred to as pancake bouncing.  
A spectacular reduction in contact time by factor of four was reported. Further experimental studies of similar macro-textures can be found in \cite{Liu2015NatCom,Gauthier2015}. 

Although pancake bouncing was shown to reduce the contact time significantly, questions remain regarding the physics behind the phenomenon including the role played by surface energy, viscous dissipation and influence of air pockets that might be trapped between the droplet and the surface texture. 
Also a parametric study including the dependence on the texture geometry, quality of coating, velocity of the drop etc. can help understanding the limits and optimizations of the macro-texture proposed in Ref.\ \cite{liu2014pancake}.

Such detailed analysis and information regarding the complex droplet shape and its deformation can be made possible through simulations. 
To that end, simulations of Ref.\ \cite{moevius2014pancake} were able to capture pancake bouncing in a qualitative manner. 
However, due to limitations of the lattice Boltzmann model used in Ref.\ \cite{moevius2014pancake}, a quantitative comparison was restrictive.
First, only a quasi-three-dimensional simulation were performed (cylindrical droplet instead of spherical) and only square posts rather than the tapered posts were considered. Another limitation was due to the high relative density of the gas phase which precluded the study of the actual surface geometry.

In this paper, we report a comprehensive simulation study of the pancake bouncing effect  and outline a new energy analysis technique that could reveal the interplay between kinetic, surface and viscous forces that influence droplet wall interactions. The recently introduced entropic lattice Boltzmann method (ELBM) for two-phase flows \cite{PhysRevLett.114.174502} is free of the aforementioned limitations and enables us to consider complex texture with the realistic geometry.   
First, the validity and the accuracy of the numerical simulations is established by comparing the simulation results with those recently observed by experiments for pancake bouncing on superhydrophobic surfaces in Ref \cite{liu2014pancake} and interaction of a droplet with a flat surface. Then a detailed parametric study is conducted by varying the geometry of the surface and also the contact angle of the substrate. Interesting analysis of various forces and energies acting during the collision process is also provided. It is important to note that the model used here is free of an tuning parameters and case based modeling. The simulation algorithm including the liquid-vapor interactions and the fluid-solid interactions remain the same for collision of two droplets and also collision of a droplet with a flat or complex wall \cite{Mazloomi2015PRE,PhysRevLett.114.174502,Mazloomi2016PHF}. Such accurate and reliable simulations combined with novel analysis techniques can uncover the physics behind these droplet wall interactions and lead to the design, optimization and also discovery of new surfaces.

The outline of the paper is as follows: In sec.\ \ref{sec:LBM}, we briefly explain our numerical model. 
In second section, we present the geometry of simulations and the results obtained by ELBM simulations are compared
with experimental observations of Ref \cite{liu2014pancake}. After that, new findings regarding the dynamics of the pancake bouncing for a drop impacting a surface with tapered posts, are presented. Finally, the paper is summarized in last section.

\section{Simulation method}
\label{sec:LBM}

We use the entropic lattice Boltzmann model (ELBM) for two-phase flows \cite{PhysRevLett.114.174502}. 
The method was discussed in detail elsewhere \cite{PhysRevLett.114.174502,Mazloomi2015PRE,Mazloomi2016PHF}; a brief summary is given below. ELBM equation for the populations $f_i(\bm{x},t)$ of the discrete velocities $\bm{v}_i$, $i=1,\dots,N$, reads,
\begin{align}
\begin{split}
f_i(\bm{x}+\bm{v}_i\delta t,t+\delta t)= f_i(\bm{x},t)+
\alpha \beta\left[f_i^{\rm eq}(\rho,\bm{u})-f_i(\bm{x},t)\right]
+ \\
[f_i^{\rm eq}(\rho,\bm{u}+\delta \bm{u}) - f_i^{\rm eq}(\rho,\bm{u})], 
\label{eq:Collision}
\end{split} 
\end{align}
We use the lattice with $N=27$ discrete velocities $\bm{v}_i=(v_{ix},v_{iy},v_{iz})$ where $v_{i\xi}=\{\pm 1,0\}$. 
The equilibrium populations $f_i^{\rm eq}$ minimize the discrete entropy function $H=\sum_{i=1}^N f_i\ln(f_i/W_i)$ under fixed density and momentum, $\{\rho,\rho \bm{u}\}=\sum_{i=1}^N\{1,\bm{v}_i\}\{f_i^{\rm eq}\}$, where $W_i$ are the lattice weights \cite{karlin1999,ansumali2003minimal,chikatamarla2006entropic}.
Parameter $0 < \beta < 1$ is fixed by the kinematic viscosity $\nu$ through $\nu=c_{\rm s}^2\delta t[1/(2\beta)-1/2)]$.
Here $c_s=\delta x/(\sqrt{3}\delta t)$ is the lattice speed of sound; lattice units $\delta x=\delta t=1$ are used. 
The relaxation parameter $\alpha$ is computed at each lattice site and at every time step from the entropy balance condition \cite{karlin1999}. The latter provides numerical stability without compromising on the accuracy thus significantly reducing the grid requirements for the simulation at high Weber and Reynolds numbers. 

Furthermore, in (\ref{eq:Collision}), the last term implements the phase separation and fluid-solid interaction through evaluation of the flow velocity increment, $\delta\bm{u}=(\bm{F}/\rho)\delta t$, with the force $\bm{F}=\bm{F}_{\rm f}+\bm{F}_{\rm s}$. The mean-field force $\bm{F}_{\rm f}=\nabla\cdot \left(\rho c_{\rm s}^2 \bm{I}-\bm{P}\right)$ implements Korteweg's stress  \cite{Korteweg1901,Rowlinson},
\begin{equation}
\bm{P}= \left(p-\kappa \rho \nabla ^2\rho-\frac{\kappa}{2}\left|\nabla \rho \right|^2\right)\bm{I}+\kappa (\nabla\rho) \otimes (\nabla\rho),
\label{eq:Pabtarget}
\end{equation}
where $\kappa$ is the coefficient which controls the surface tension,  $\bm{I}$ is unit tensor and $p$ is the equation of state; the Peng-Robinson form is used for the latter \cite{PhysRevLett.114.174502,Mazloomi2015PRE,Mazloomi2016PHF}. 
%
%
Interaction between the fluid and the solid surface is introduced with the help of a force $\bm{F}_{\rm{s}}$,
\begin{equation}
\bm{F}_{\rm{s}}(\bm{x},t)=\kappa_{\rm w} \rho(\bm{x},t) \sum_{i} w_{i}s(\bm{x}+\bm{v}_{i}\delta t)\bm{v}_{i},
\label{f-s:forces}
\end{equation}
where parameter $\kappa_{\rm w}$ reflects the intensity of the fluid-solid interaction. 
By adjusting $\kappa_{\rm w}$, solid surfaces with different wetting can be modeled.  
In Eq.\ (\ref{f-s:forces}),  $s(\bm{x}+\bm{v}_{i}\delta t)$ is an indicator function that is equal to one for the solid domain nodes, and is equal to zero otherwise; $w_i$ are appropriately chosen weights \cite{yuan2006,Mazloomi2015PRE}. 
\begin{figure}
\centering
\includegraphics[width=0.75\textwidth]{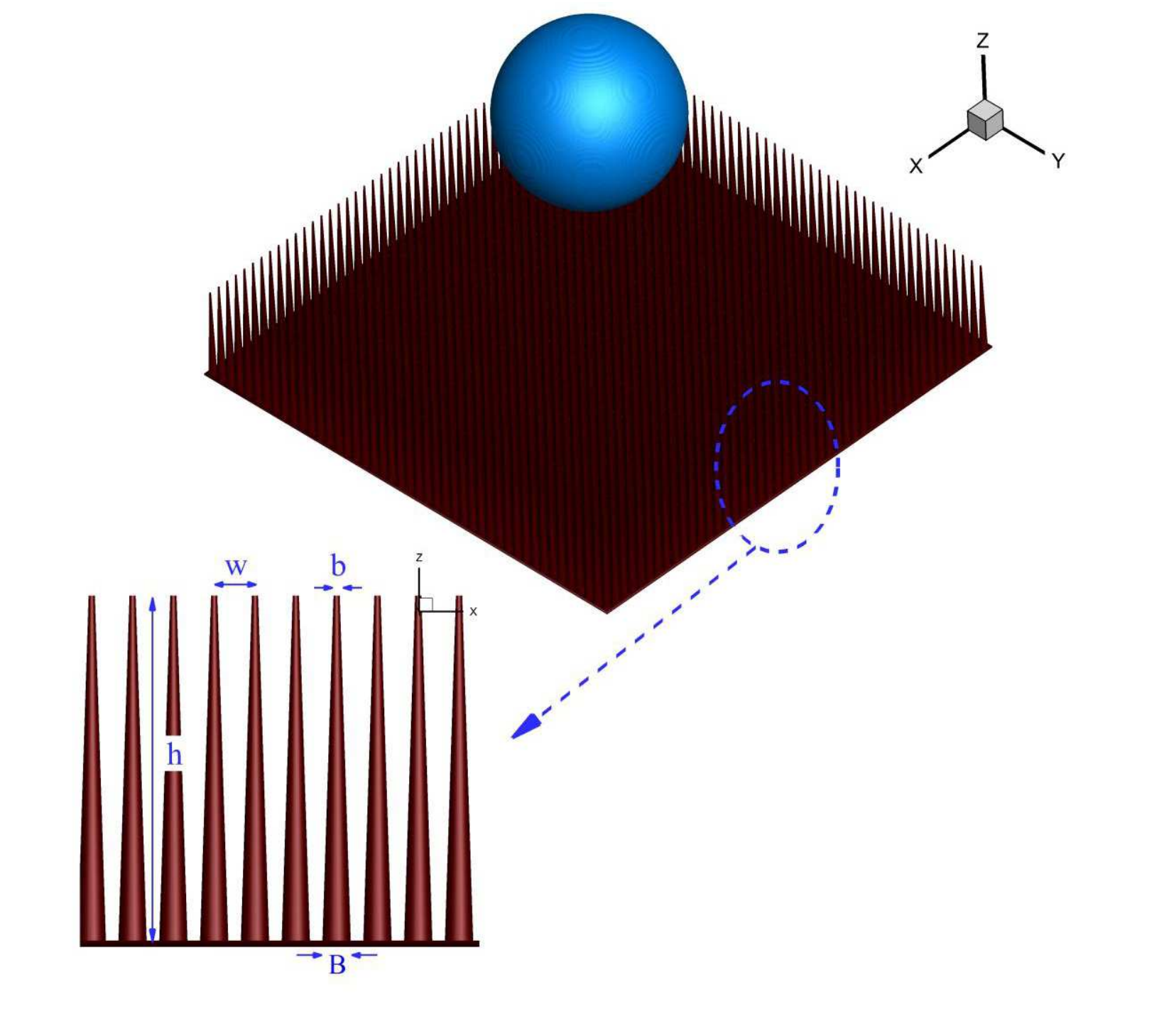}
\caption{Simulation setup. The texture is represented a surface decorated with a lattice of tapered posts. 
	The posts are represented by right conical frustums with the base diameters $b$ and $B$; $w$ is the center-to-center spacing and $h$ is the height of the posts.}
\label{fig:Schematic}
\end{figure}
\section{Results and discussion}
\label{sec:Results}
\subsection{Geometry and simulation parameters}

The setup of three-dimensional simulations is sketched in Fig.\ \ref{fig:Schematic}. 
A droplet of radius $R_0$ is placed above the surface in equilibrium with the vapor. A uniform downward velocity is imposed on the drop while surrounding vapor is initialized with zero velocity. Initially the drop is sufficiently elevated to allow the liquid-vapor interface to equilibrate before the impact. No gravity is considered in the simulations due to the small time scales involved. All simulations, unless otherwise stated, were run on computational domain of $6R_0\times6R_0\times6R_0$.
Parameters of the simulated fluid are the same as the ones employed for investigation of droplets collisions \cite{Mazloomi2016PHF} and the droplet-surface interactions \cite{Mazloomi2015PRE}:  $\rho_l=7.82$ (liquid density), $\rho_v=0.071$ (vapor density), $\sigma=0.353$ (surface tension, corresponding to $\kappa =0.00468$ in Eq.(\ref{eq:Pabtarget})) and $\mu_v=0.01$ (vapor dynamic viscosity); all are in lattice units (see below the match to physical units). The impact velocity was computed in accord with the Weber number  ${\rm We}=\rho_l R_0 U_0^2/\sigma$, and the range $6 \leq {\rm We} \leq 150$ was studied. The dynamic viscosity of the droplet was fixed at  $\mu_l=0.415$ corresponding to Ohnesorge number ${\rm Oh}=0.025$ (${\rm Oh}=\mu_l/\sqrt{\rho_l\sigma R_0}$). This corresponds to a range of Reynolds number ${\rm Re}=\sqrt{{\rm We}}/{\rm Oh}$ as $98\le {\rm Re}\leq 490$. A comment on the parameters of the simulated liquid and those of the used in the experiment \cite{liu2014pancake} (water) is in order. The density contrast $\rho_{\rm l}/\rho_{\rm v}\approx 100$ and the Ohnesorge number ${\rm Oh}=0.025$ used in the ELBM simulations were shown before to be sufficient to recover the pertinent flow dynamics of colliding droplets and impacts on flat surfaces \cite{Mazloomi2015PRE,Mazloomi2016PHF}. The density contrast for water is an order of magnitude higher, whereas the Ohnesorge number is an order of magnitude lower than those of the simulation (${\rm Oh}\approx 0.003$ for water). Nevertheless, similar to previously studied cases we expect the essential dynamics to be captured correctly also in the present simulations with complex textures since the vapor state is too light to influence the liquid \cite{Mazloomi2015PRE,Mazloomi2016PHF}. Further evidence is provided below by comparing ELBM simulations with experiments. 

The structure of the substrate is a square lattice of tapered posts placed uniformly with the center-to-center spacing $w$ on a flat plate. Each post is modeled as a right conical frustum of the height $h$, with the smaller and larger base diameters $b$ and 
$B$, respectively. The droplet radius $R_0$ is the input of the simulations while the rest of the geometry matches the experiment \cite{liu2014pancake}: $R_0/h=1.8$ (droplet radius to post height), $b/B \approx 0.28$, and $b=B-2h\tan\varphi$ ($\varphi=2.6^\circ$ is the apex angle \cite{liu2014pancake}).
We introduce the density of the texture (DoT) $\Lambda=R_0/w$ (parameter $\Lambda$ reflects the relative number of posts seen by the droplet at impact).

The spacing between the posts was chosen to reproduce $\Lambda_{\rm exp}\approx 7.25$ \cite{liu2014pancake} in the simulations of sec.\ \ref{sec:comparison} below. 
Both the surface of the conical frustum and the supporting flat plate are considered SHS with a contact angle $\theta=165^\circ$. 

Finally, in order to convert lattice time $t_{\rm LB}$ into seconds, we first compute the inertia-capillary time $\tau_{\rm LB}=\sqrt{\rho_{\rm l} R_0^3/\sigma}$ using the density, droplet radius and surface tension in lattice units. Next, $\tau$ is extracted from the experimental data and the reduced time for both the experiment and the simulation are matched, $t_{\rm LB}/\tau_{\rm LB}=t/\tau$. Thus, given $t_{\rm LB}$ (the number of time steps), we uniquely obtain the corresponding physical time $t=(\tau_{\rm LB}/\tau) t_{\rm LB}$. 
A droplet radius of $R_0=100$ grid units was used unless stated otherwise. All simulations were checked for grid convergence using $R_0=100$ and lower; only highest resolution results are reported here. 

\subsection{Validation with experimental data}
\label{sec:comparison}

\begin{figure}
\centering
	\includegraphics[width=0.65\textwidth]{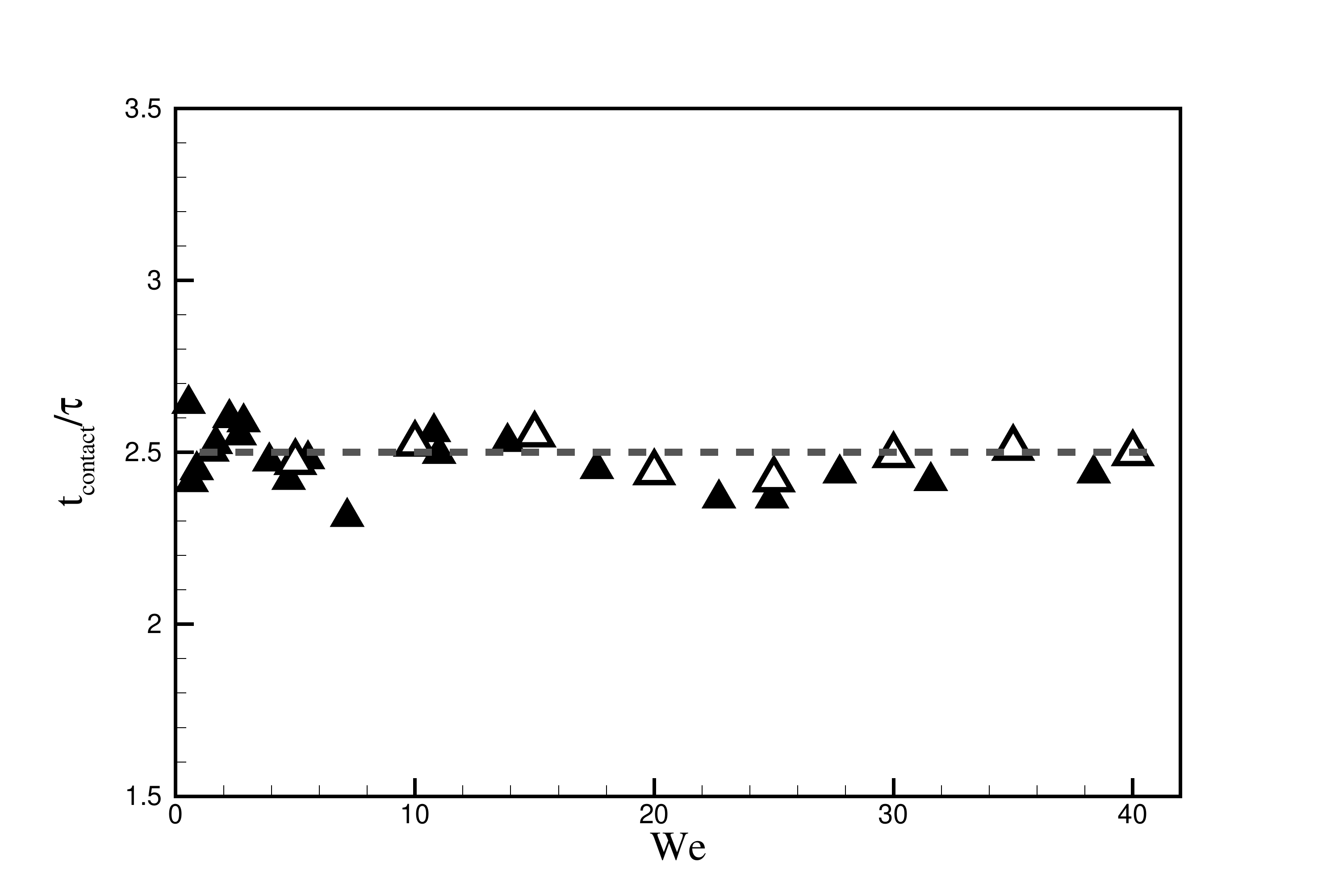}
	\caption{{Contact time on a flat super-hydrophobic surface with the contact angle of $161^{o}$ as a function of Weber number. Dashed line: $t^*_{\rm contact} = 2.5 $, Ref.\ \cite{richard2002surface}; Open symbols: ELBM simulations, solid symbols: Experiment \cite{gauthier2015water}.}}
	\label{fig:ContatTimeFlat}
\end{figure}

In this section we validate the simulations with published data from experiments \cite{gauthier2015water,liu2014pancake}.
For a flat SHS, it was shown in Ref.\ \cite{richard2002surface} that the contact time $t_{\rm contact}$ is independent of Weber number in a wide range and can be scaled with the inertial-capillary timescale: $t^*_{\rm contact}=t_{\rm contact}/\tau \approx 2.2$. 
Figure \ref{fig:ContatTimeFlat} reports the variation of contact time on the flat SHS for a range of Weber numbers. From Fig.\ \ref{fig:ContatTimeFlat} the results obtained from simulations (open symbols) agree well with those observed by experiment (solid symbol) in \cite{gauthier2015water}.  


We note that the flat SHS test validates the use of the Ohnesorge number ${\rm Oh}=0.025$ for the ELBM fluid in the present context. Indeed, as it has been shown by many authors, e.\ g.\ \cite{Antonini2013,Maitra2014}, the Ohnesorge number takes effect on the contact time at much higher values, ${\rm Oh}\approx 1$ (i.\ e.\ for highly viscous liquids such as glycerol (${\rm Oh}\approx 3$) or silicon oil). For ${\rm Oh}<0.1$, there is no effect of viscosity during impact on flat surfaces, and thus it is not surprising that the present simulation agrees well with the experiments which use water. We refer to \cite{Mazloomi2015PRE} for other comparisons of ELBM with experiments on flat SHS.

\begin{figure}
\centering

\subfigure[]{
   \includegraphics[width=0.85\textwidth] {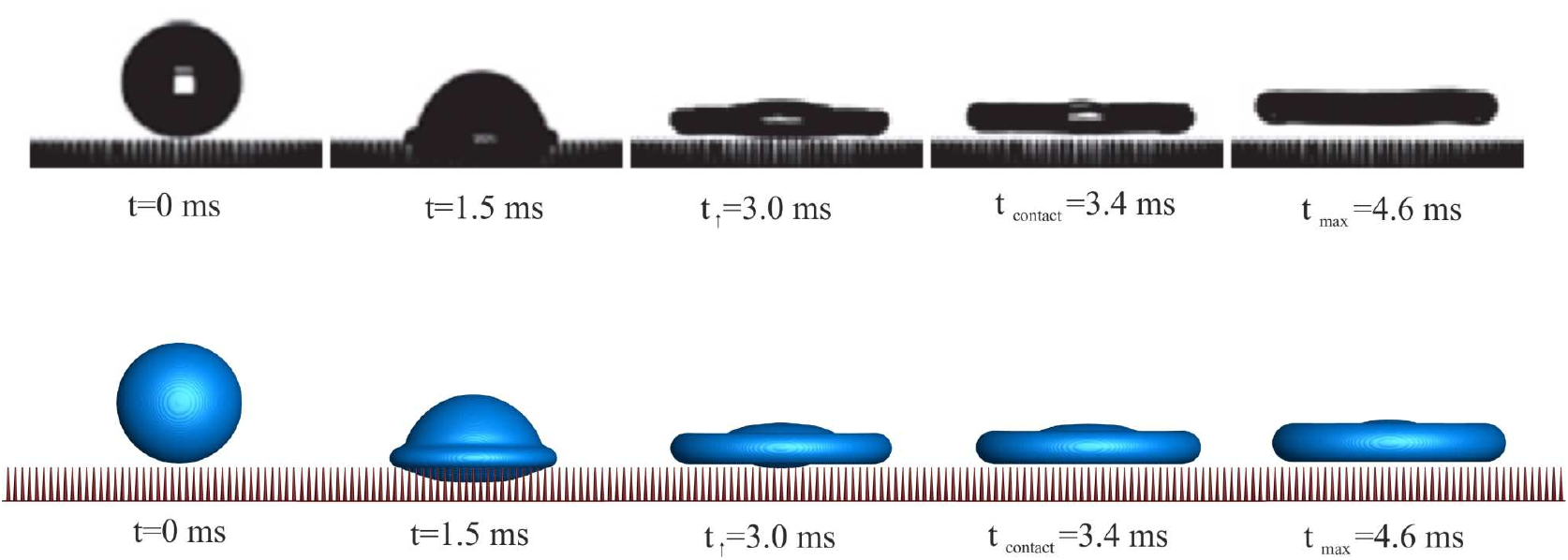}
   \label{fig:tapered}
 }

 \subfigure[]{
   \includegraphics[width=0.85\textwidth] {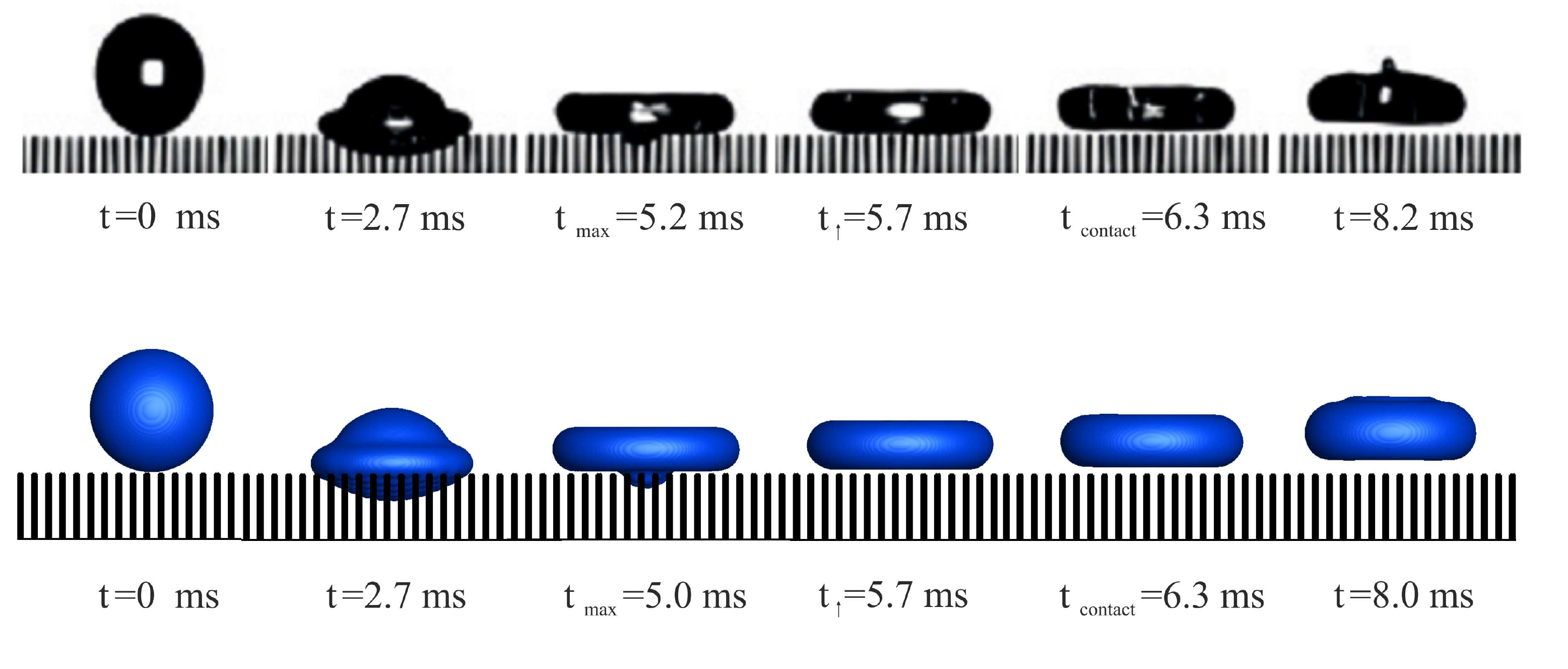}
   \label{fig:square}
 }

\label{fig:tapered_square_posts}
\caption{Comparison of simulation (bottom) and experiment \cite{liu2014pancake} (top) for the pancake bouncing of a drop impinging on, (a) the tapered posts at ${\rm We}=14.1$, (b) the straight square posts at ${\rm We}=7.9$.}
\end{figure}

Now we proceed with the textured SHS. Following Ref.\ \cite{liu2014pancake}, we introduce characteristic time instances ($t=0$ corresponds to the time of first contact): 
$t_\uparrow$ is the time at which the texture is fully emptied; at $t_{\rm max}$ the drop reaches its maximal lateral extension; $t_{\rm contact}$ is the time at which the drop looses contact with the surface.
Snapshots of a drop impinging on a texture of tapered posts at ${\rm We}=14.1$ are shown in Fig.\ \ref{fig:tapered} along with the images from the experiment \cite{liu2014pancake}. Pancake formation and rebound is clearly seen in the simulation.
Simulation results for both the shape of the drop and the characteristic times are in excellent agreement with experimental observations. 

Along with the tapered posts texture of Fig.\ \ref{fig:Schematic}, we simulated a simpler case of rectangular prism posts which was also considered in the experiment of Ref.\ \cite{liu2014pancake}. 
The droplet radius for this simulation was $R_0 = 34$ lattice units.
The height of posts $h$, the posts center-to-center spacing, $w$ and the side of the square cross-section $b$  were computed according to the aspect ratios $R_0/h$, $R_0/w$ and $b/h$ of the experiment \cite{liu2014pancake} and the results are presented in Fig.\ \ref{fig:square} for  ${\rm We}=7.9$.
Also in this case excellent agreement between simulation and experiment is observed.


\begin{figure}[ht!]
\centering

\subfigure[]{
   \includegraphics[width=0.65\textwidth] {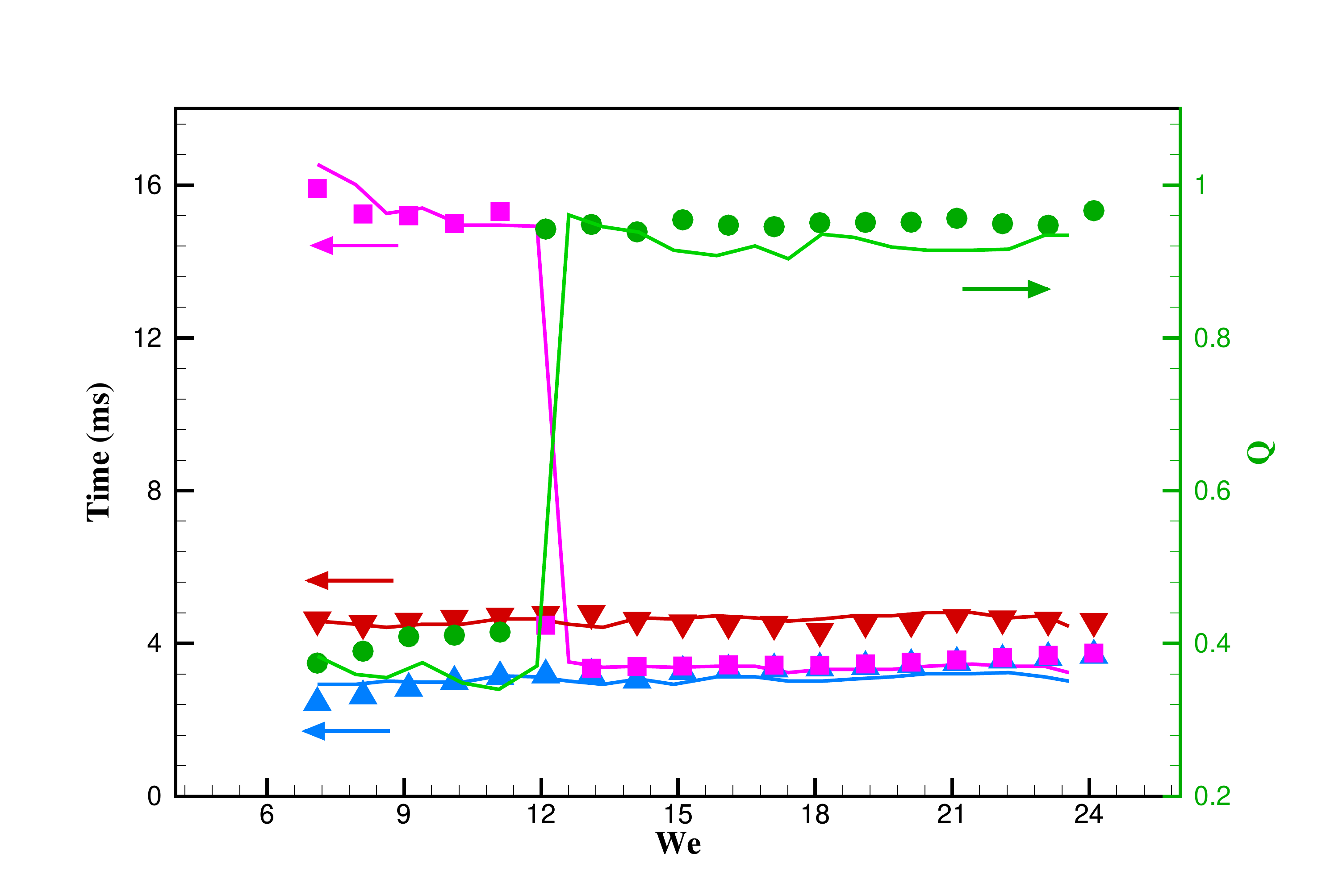}
   \label{fig:timecales_tapered}
 }

 \subfigure[]{
   \includegraphics[width=0.65\textwidth] {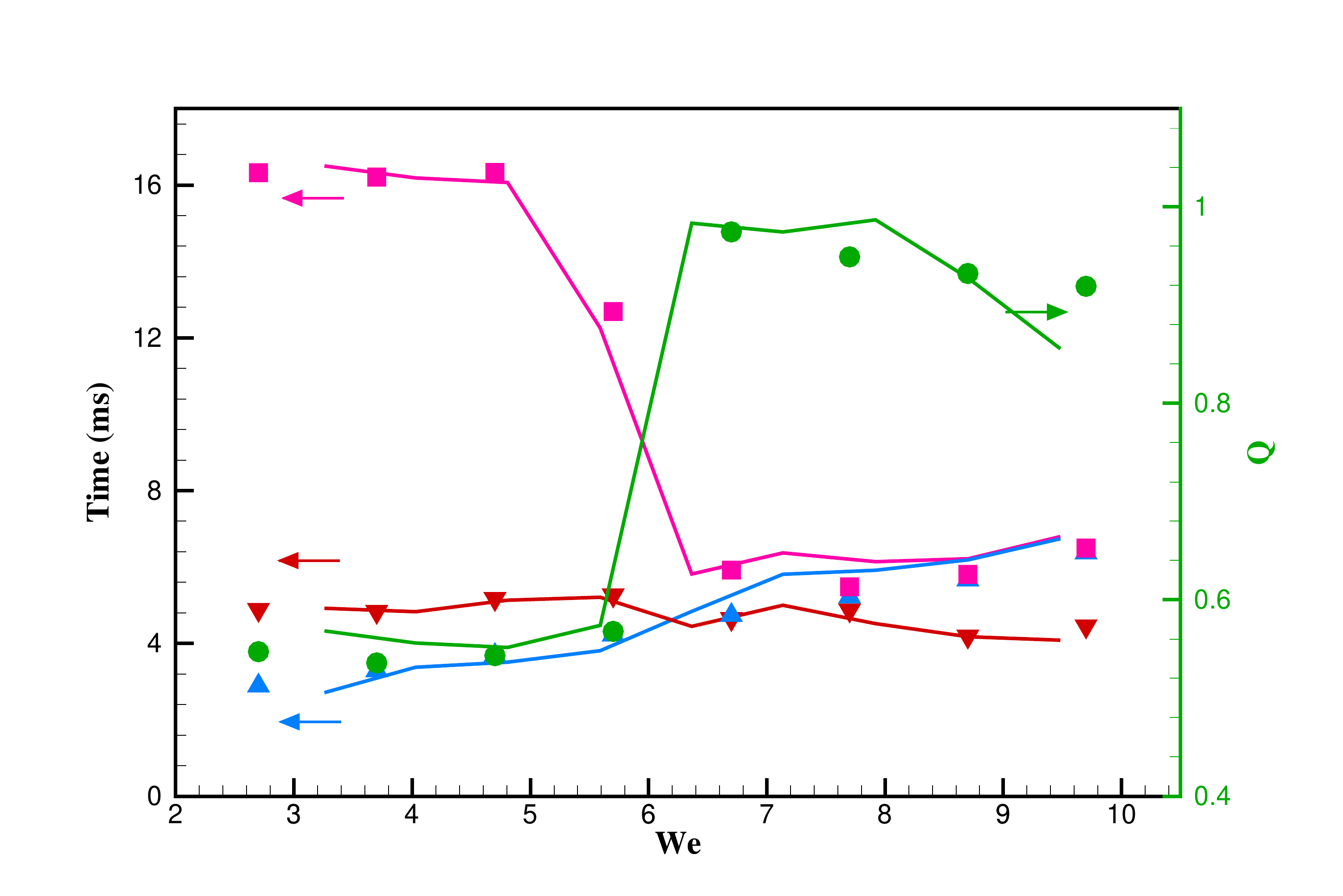}
   \label{fig:timecales_square}
 }

\label{fig:timecales}
\caption{Comparison of simulation (symbol) and experiment \cite{liu2014pancake} (line) of characteristic times  $t_\uparrow,t_{\rm max},t_{\rm contact}$, and pancake quality, $Q=d_{jump}/d_{max}$, with We for a drop impinging on, (a) tapered posts, (b) square posts. The blue, red, pink and green colors show the $t_\uparrow,t_{\rm max},t_{\rm contact}$ and $Q$, respectively }
\end{figure}
%

%
For a validation, simulations were performed within the experimentally accessed range of Weber numbers $6<{\rm We}<25$. Dependence of characteristic times $t_\uparrow$, $t_{\rm max}$, and  $t_{\rm contact}$ on Weber number is compared with the experiment in Fig.\ \ref{fig:timecales_tapered} (tapered posts) and Fig.\ \ref{fig:timecales_square} (square posts). Also the so-called pancake quality  $Q=d_{\rm jump}/d_{\rm max}$ is compared to the experiment, where  $d_{\rm jump}$ and $d_{\rm max}$ are the diameters of the drop at $t_{\rm contact}$ and $t_{\rm max}$, respectively. 


Simulations in Fig.\ \ref{fig:timecales_tapered} show that for ${\rm We}<12$ the contact time is $t_{\rm contact}\simeq16 [ms]$ which is in good agreement with the conventional complete rebound from a flat surface \cite{richard2002surface}.
The onset of the pancake bouncing regime at critical Weber number ${\rm We}^*\approx 12$ agrees well with experiment \cite{liu2014pancake}. Reduction in contact time by a factor four is observed.
We repeated the measurements of the characteristic times scales for the drop impinging on the straight square posts. 
Simulation results compare well with the experiment in Fig.\ \ref{fig:timecales_square}. 

Summarizing, ELBM simulation demonstrates excellent agreement with the existing experimental data. 
In the remainder of this paper, we shall address regimes which were so far not studied experimentally in order to gain a more comprehensive picture of bouncing off  macro-textured surfaces.


\subsection{Textures with perfect coating}
\label{PerfectSur}

\begin{figure}[ht!]
	\includegraphics[width=0.95\textwidth]{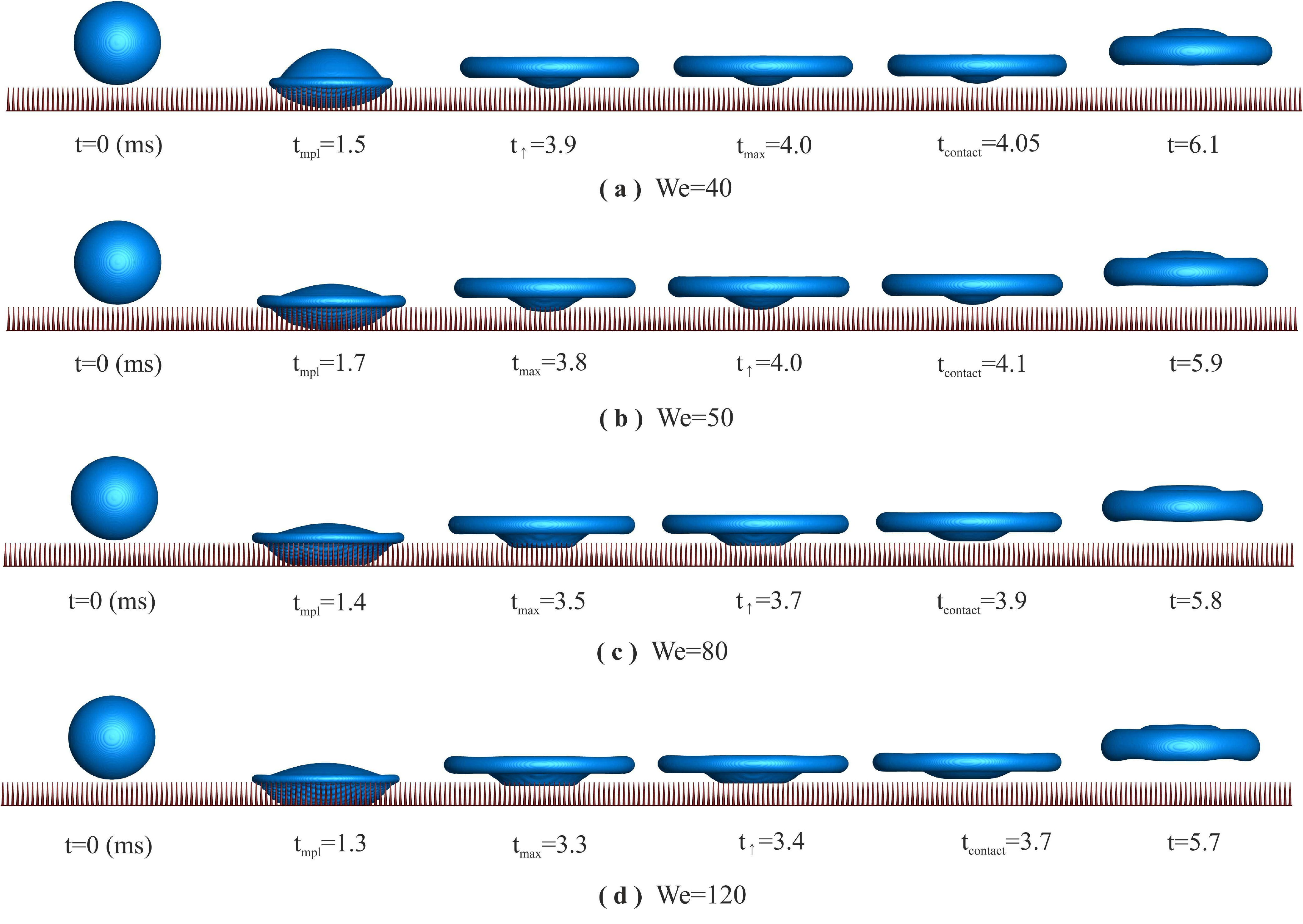}
	\caption{ Snapshots of the rebound from tapered posts at (a) ${\rm We}=40$, (b) ${\rm We}=50$, (c) ${\rm We}=80$, (d) ${\rm We}=120$. Density of the texture $\Lambda_{\rm exp}=7.25$. Invading liquid hits the base of the texture at ${\rm We}\geq 50$. Perfect coating is assumed for both the posts and the base plate (contact angle is set to $\theta=165^\circ$). After hitting the base, penetrated liquid experiences a quick lateral extension, detaches from the base, returns to the top of the posts and demonstrates pancake rebound (Supplementary Movies 4 and 5). }
	\label{fig:We40_50_80_120}
\end{figure}

\begin{figure}
\centering
	\includegraphics[width=0.65\textwidth]{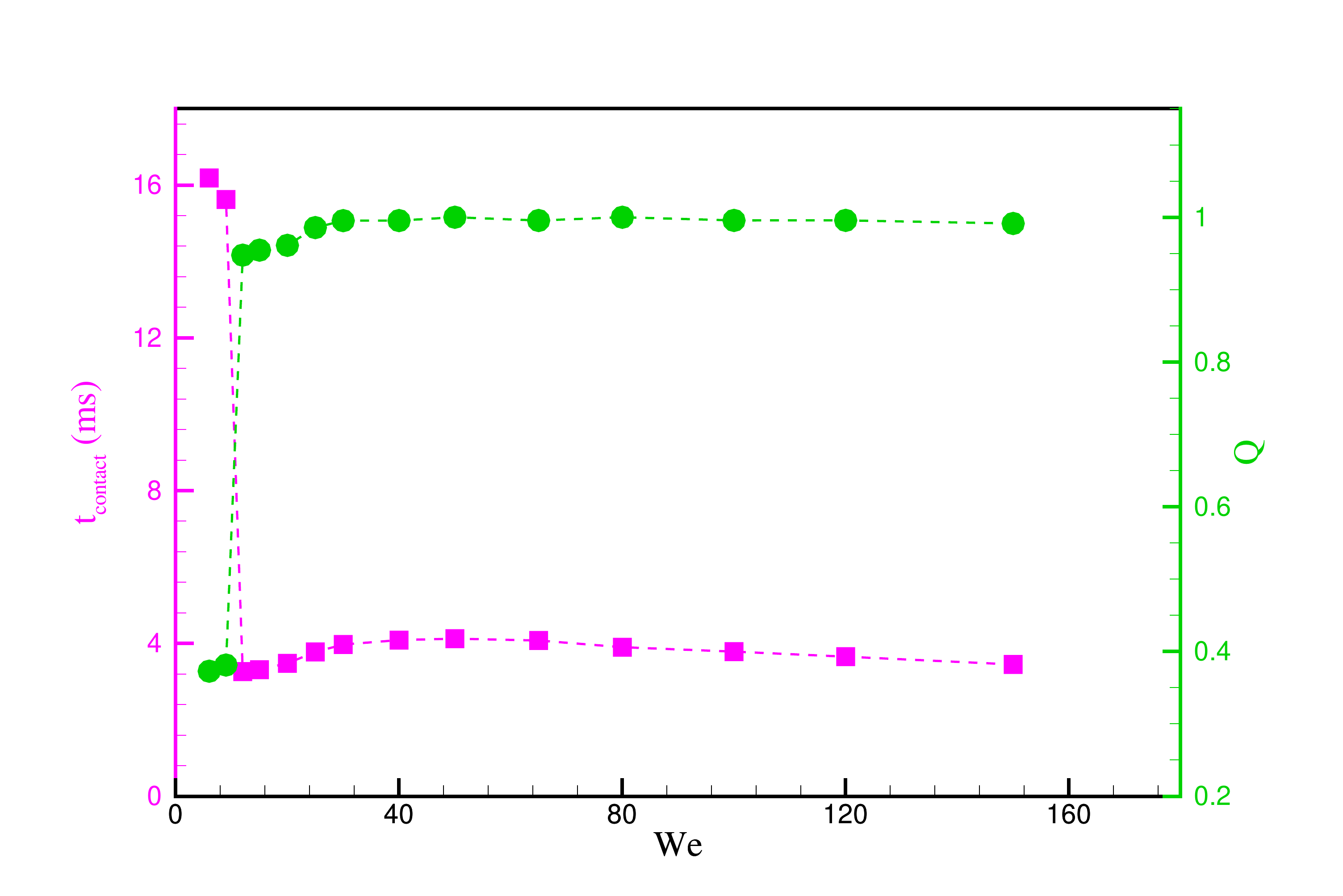}
	\caption{Contact time $t_{\rm contact}$ (squares) and pancake quality $Q$ (circles) of a drop impinging on perfectly coated tapered posts with the density of texture $\Lambda_{\rm exp}=7.25$, for a range $6\leq {\rm We}\leq 150$.}	\label{fig:tcontact}
\end{figure}

\begin{figure}
\centering
	\includegraphics[width=0.65\textwidth]{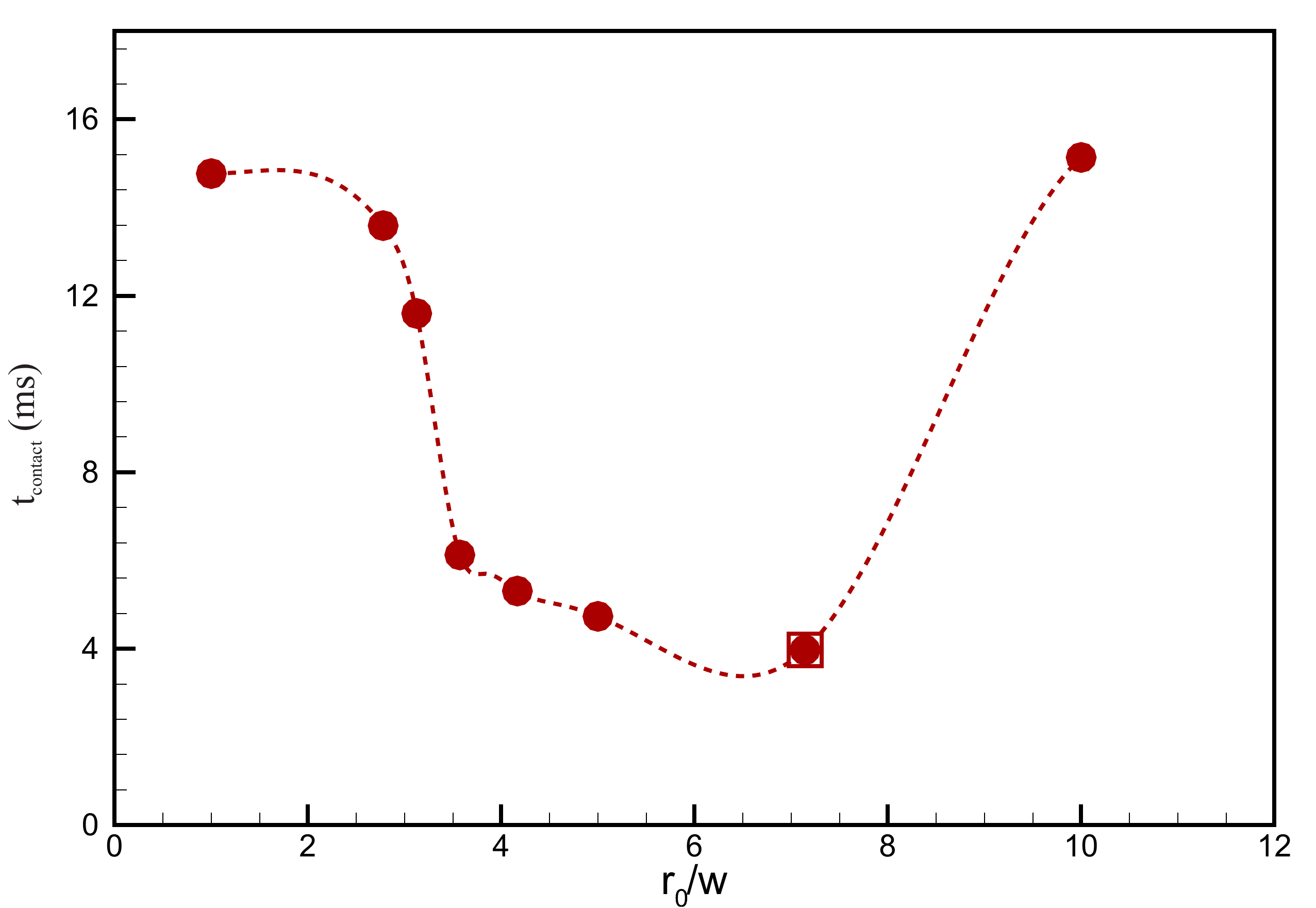}
	\caption{ Dependence of the contact time on the density of	the  texture $\Lambda=R_0/w$ at ${\rm We}=30$. DoT of the experiment \cite{liu2014pancake} is marked with a square, $\Lambda_{\rm exp}=7.25$.} 
	\label{fig:tcontact_spacing}
\end{figure}

Ref.\ \cite{liu2014pancake} reported experiments in a narrow range of Weber numbers (${\rm We}<25$). When the Weber number is increased,  penetration depth into the texture increases and the deforming drop will eventually reach the base of the substrate. It is difficult to study the liquid inside the texture in the experiments. 
Here we extend the ELBM simulations to higher Weber numbers, $6<{\rm We}<150$. We first consider the same geometry as in the previous section, and assume perfect coating, that is, both the posts and the base are SHS with contact angle $\theta=165^\circ$.

Fig.\ \ref{fig:We40_50_80_120} shows snapshots of the impact on the perfectly coated texture at various Weber numbers. 
For ${\rm We} > 40$, the penetrated liquid interacts with the base plate at the time $t_{\rm mpl}$ (maximal penetration of the drop). 
Simulations show that pancake bouncing is still observed for a much wider range of Weber numbers, $12 < {\rm{ We}} < 150$. 
Fig.\ref{fig:tcontact} shows that both the reduction of the contact time (squares)  and the pancake quality (circles) remain unaffected until at least ${\rm We}\approx150$. This can be attributed to the fact that the base plate of the texture is also considered SHS with a uniform contact angle (see sec. \ref{sec:coating}). 


Next we numerically probe the effect of lower and higher lateral density of the texture by varying the center-to-center spacing.
Contact time for various $R_0/w$ is shown in Fig.\ \ref{fig:tcontact_spacing} for a selected Weber number ${\rm We}=30$. For a denser texture ($R_0/w \approx 10$), since the drop requires larger kinetic energy to start penetrating into the posts, only a small portion of the liquid drop penetrates into posts and thus adequate capillary energy required for pancake bouncing is not stored by the penetrated liquid. Therefore, the drop impinging on such a dense texture follows a conventional rebound pattern. 
On the other hand, when the tapered posts are distributed sparsely ($R_0/w < 7.25$), the drop can penetrate the texture at a lower critical Weber number. However, in this case the total interface area and hence surface energy stored for a given droplet penetration is lower due to a less proliferated interface. 
When a drop of a given radius meets the texture with sparsely distributed posts, lesser number of penetration valleys (undulated regions) are produced. This results in lower surface energy and lower capillary forces during the emptying process; thus the contact time increases.   
Hence, during a design process, one needs to consider the trade off between contact time reduction and critical Weber number at which the process of pancake bouncing sets in. 

Summarizing, under the assumption of perfect SHS coating, we found that the pancake bouncing extends to much higher Weber numbers whereas the density of the texture strongly affects the critical Weber number. In the next section we shall investigate a more realistic scenario of imperfect coating.

\subsection{Imperfect coating}\label{sec:coating}

In the simulations so far, we assumed that the SHS quality is maintained uniformly throughout the texture and the base plate. 
However, this is unlikely to hold when, for example, the  SHS coating is produced by spraying a polymer solution on the texture. 
According to Ref.\ \cite{liu2014pancake}, controlling the quality and uniformity of coating throughout the the posts and especially the valleys between them is a difficult task. 
Hence it is reasonable to assume that the contact angle at the base plate is lower than that on the upper part of the posts. 
Now, given that the drop interacts with the base plate at ${\rm We} \geq 50$ (see Fig.\ \ref{fig:We40_50_80_120}), we probe the effect of imperfect coating by assigning a smaller contact angle of $\theta_{\rm{bottom}}=140^\circ$ to the base plate and the bottom part of the posts ($10\%$ of the height). Since superhydrophobic surfaces are created by coating hydrophobic surfaces with a nono-scale coating, this assumption of $\theta_{\rm{bottom}} =140$ (regular hydrophobic surface) is reasonable. The rest of the posts are maintained at the contact angle $\theta=165^\circ$.

\begin{figure}
\centering
	\includegraphics[width=0.65\textwidth]{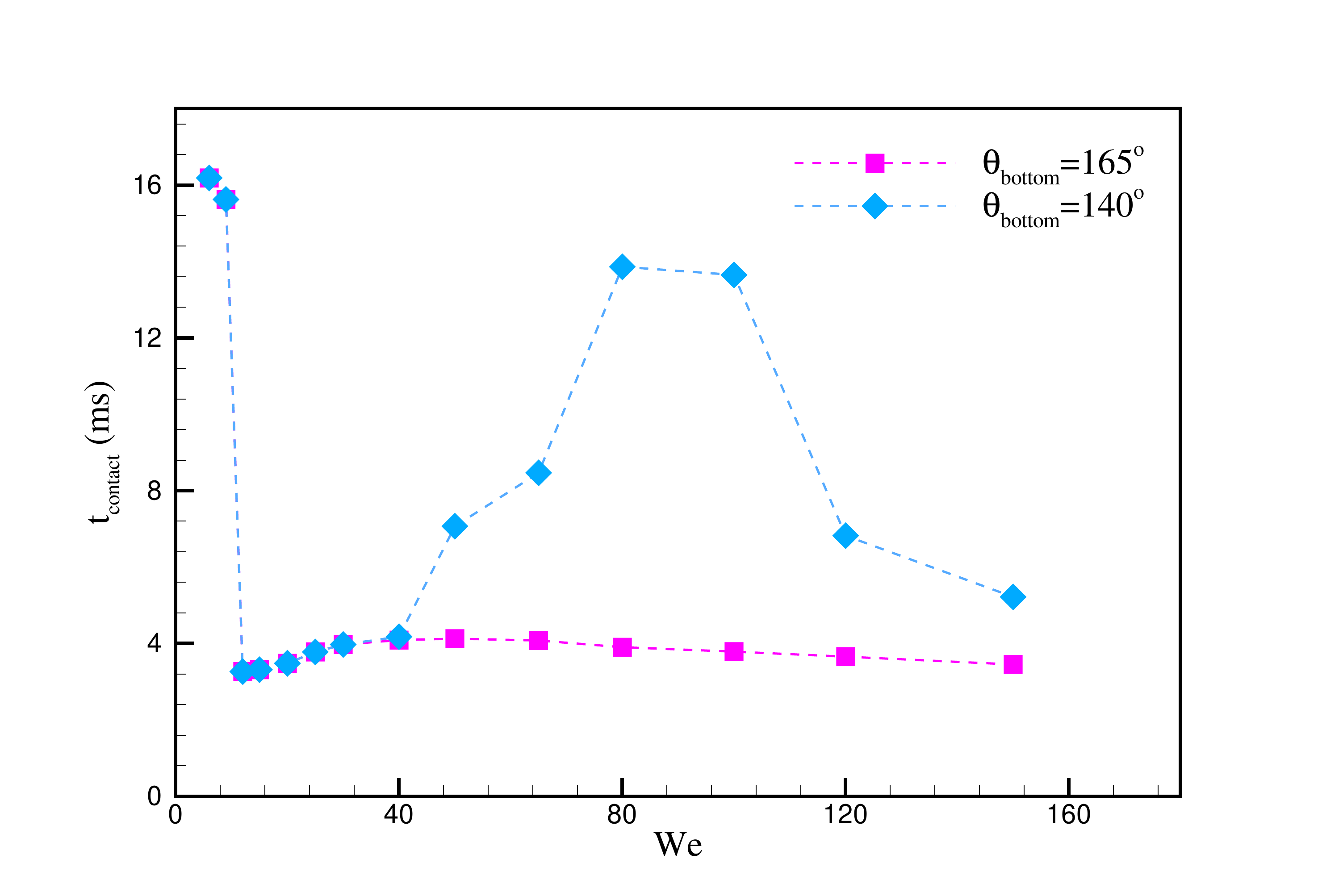}
	\caption{ Contact time $t_{\rm contact}\ [ms]$ for a perfectly (squares) and imperfectly (diamonds) coated texture as a function of Weber number.}
	\label{fig:ContactTime_BottomEffect}
\end{figure}

\begin{figure}
\centering
	\includegraphics[width=0.95\textwidth]{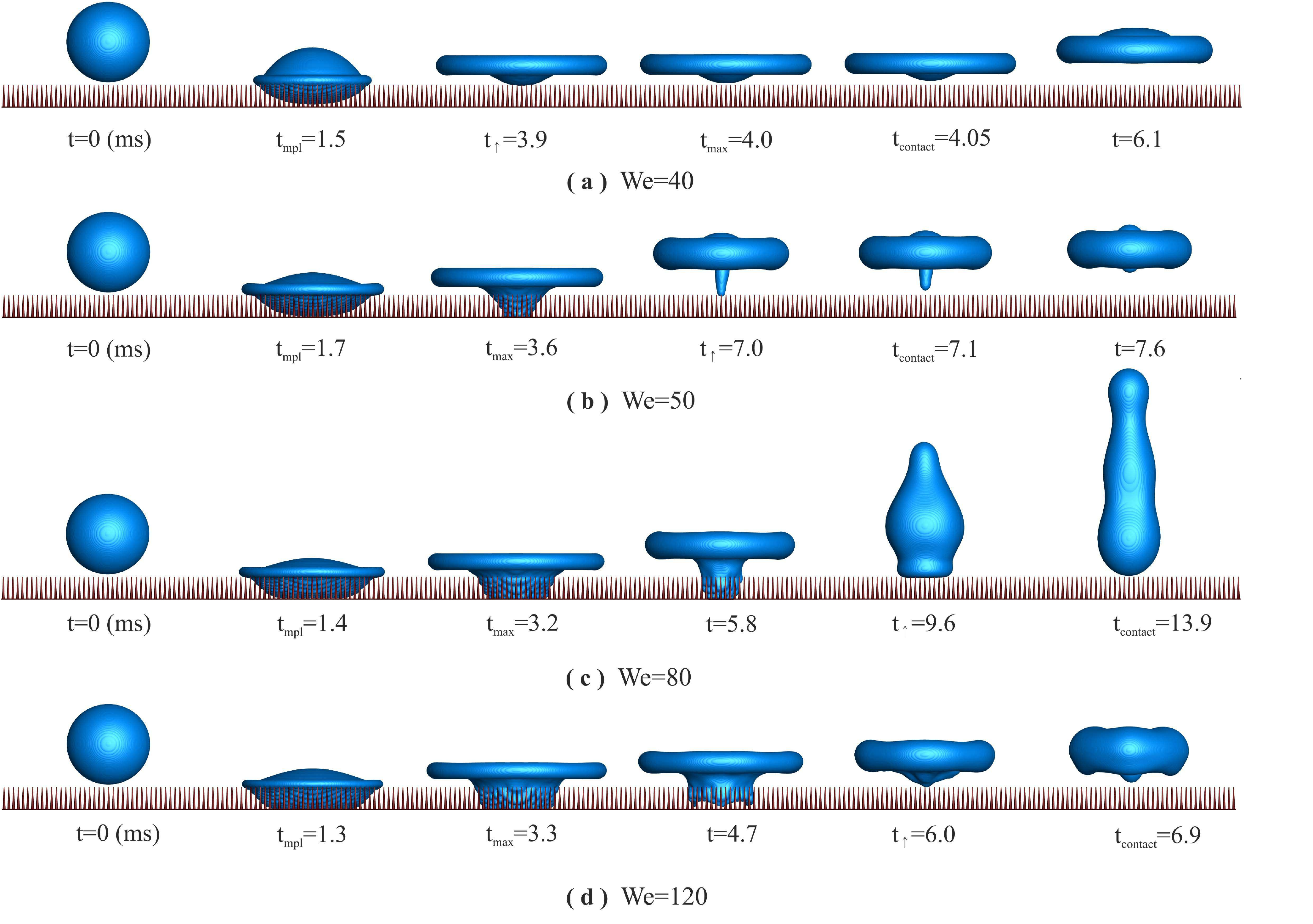}
	\caption{Snapshots of impact on imperfectly coated texture at (a) ${\rm We}=40$, (b) ${\rm We}=50$, (c) ${\rm We}=80$, (d) ${\rm We}=120$. 
		Contact angle at the base plate and $10 \%$ above it set to $\theta_{\rm bottom}=140^\circ$, and for the rest of the texture the contact angle is $\theta=165^\circ$. 
		For ${\rm We} \geq 50$, the penetrated liquid reaches the base at $t_{\rm mpl}$. (Supplementary Movie 6 and 7).}
	\label{fig:We40_50_80_120_ThetaBottom_140}
\end{figure}
 
The contact time with the perfectly and the imperfectly coated textures is shown in Fig.\ \ref{fig:ContactTime_BottomEffect}. 
For ${\rm We} \leq 40$, the contact time is the same for both textures. 
In fact, for ${\rm We} \leq 40$, since the liquid penetrates the texture without touching the base, the quality of coating does not affect the contact time. 
However, for ${\rm We} > 40$, the liquid reaches the substrate base, and the degraded coating  alters significantly the contact time droplet dynamics and the pancake quality.\\
To elucidate this, we present snapshots of the impact on an imperfect surface for different Weber numbers in Fig.\ \ref{fig:We40_50_80_120_ThetaBottom_140}. 
For ${\rm We}=50$, $80$ and $120$, the drop contacts with the substrate base at $t_{\rm mpl}$. Due to lack of super-hydrophobicity at the bottom, the penetrated liquid tends to stick to the texture base before it is pulled out by the rest of drop moving upward. 
In Fig.\ \ref{fig:We40_50_80_120_ThetaBottom_140}(b) (${\rm We}=50$), although the penetrated liquid sticks to the substrate base, it returns to the top of the texture quickly thus enabling pancake bouncing but with the almost twice higher contact time, $t_{\rm contact}\approx 7\ [ms]$. 
At a higher Weber number ${\rm We}=80$, Fig.\ \ref{fig:We40_50_80_120_ThetaBottom_140}(c), due to a larger contact area between the liquid and the texture base, the penetrated liquid returns to the top of the posts  with a delay. Consequently, the drop has enough time to retract and the overall picture is resembling the conventional bouncing rather than a pancake rebound. The contact time $t_{\rm contact}\approx 14\  [ms]$ becomes closer to the conventional bouncing value  (see Supplementary Movie 6). \\
Interestingly however, as the Weber number is further increased (see Fig.\ \ref{fig:We40_50_80_120_ThetaBottom_140}(d), ${\rm We}=120$), the contact time reduces back to the value $t_{\rm contact}\approx 7\  [ms]$. Since the contact area between the drop and the base becomes even larger, also the number of invaded valleys is larger than at ${\rm We}=80$. The force due to surface tension is thus able to overcome the pinning effect of the imperfectly coated base, and the texture is emptied faster. This explains the return of a pancake-like bouncing at ${\rm We}=120$, and the contact time becomes smaller than at 
${\rm We}=80$.  These results rely on the condition that $\theta_{\rm {bottom}} = 140$, however if the contact angle at the bottom plate is more severely effected then the deterioration in contact time is more pronounced. Due to a lack of experimental access to these bottom regions, one can only perform such inverse analysis to estimate the quality of coating in these hidden regions.\\
Summarizing, the imperfect coating at the bottom of the texture features a non-linear dependence of the contact time on the Weber number and can significantly affect the rebound pattern. 

\subsection{Energy budget}
\label{sec:energy}

Energy considerations were invoked in Ref.\ \cite{liu2014pancake} to quantify the mechanism of pancake bouncing. 
The assumption behind this analysis was that the kinetic energy of the drop is fully converted into the surface energy at the maximal penetration into the texture. However, neglecting energy dissipation is less obvious for an impact on textured surfaces. Indeed, since the shape of the droplet is considerably more distorted as compared to the flat SHS (the flow 'sees more walls'), stresses in the boundary layers contribute more to the dissipation. 
On the other hand, ELBM was shown to quantitatively capture the energy budget in binary droplet collisions \cite{Mazloomi2016PHF}. 
In this section we report and discuss the energy budget of the pancake bouncing regime from the ELBM simulations.
%
For the drop with volume $V$ and surface area $A$, let us introduce the kinetic energy, $ K=\int_{V} \frac{1}{2}\rho u^2 dV $,  the surface energy $ S = \sigma A $, and the energy loss due to viscous dissipation $\Xi$,
\begin{equation}
\Xi = \int_{0}^t\Phi dt \quad \rm{where} \quad 
\Phi = \frac{\mu_l}{2}\int_V \left(\nabla\bm{u}+\nabla\bm{u}^\dagger\right)^2dV.
\label{E_phi}
\end{equation}
Further introducing normalized energies, $\tilde{K}=K/E_0$, $\tilde{S}=S/E_0$ and $\tilde{\Xi}=\Xi/E_0$, where $E_0=K_0+S_0$ is the energy of the drop at $t=0$, energy balance is written as,
\begin{equation}
 \tilde{K} + \tilde{S} + \tilde{\Xi} =1.
\label{eq:energyconserv}
\end{equation}

\begin{figure}
	\centering
		\subfigure[$We=15$]{ %
			\includegraphics[width=0.4\textwidth] {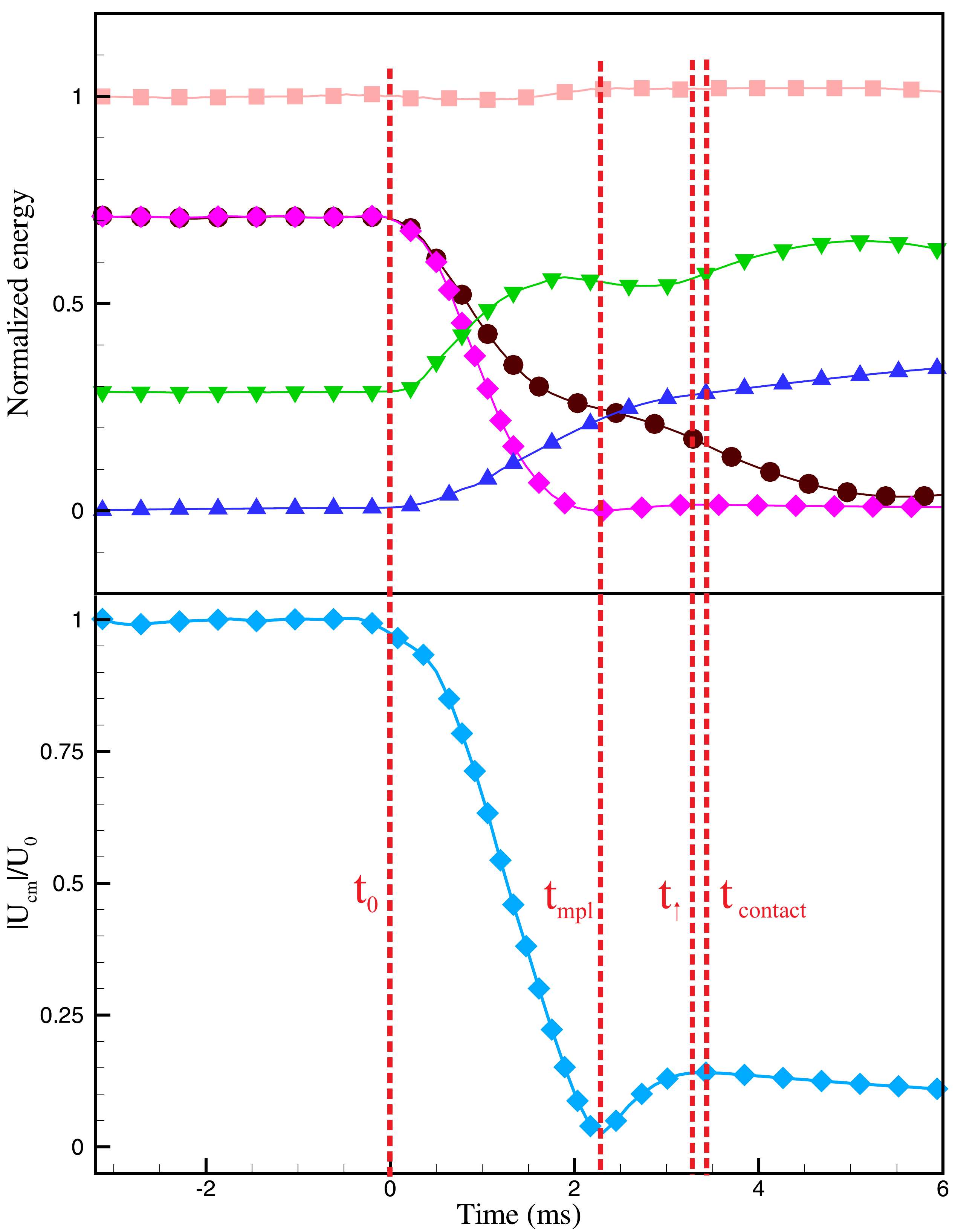}
			\label{figa}}
		\quad
	\subfigure[$We=30$]{ %
		\includegraphics[width=0.4\textwidth] {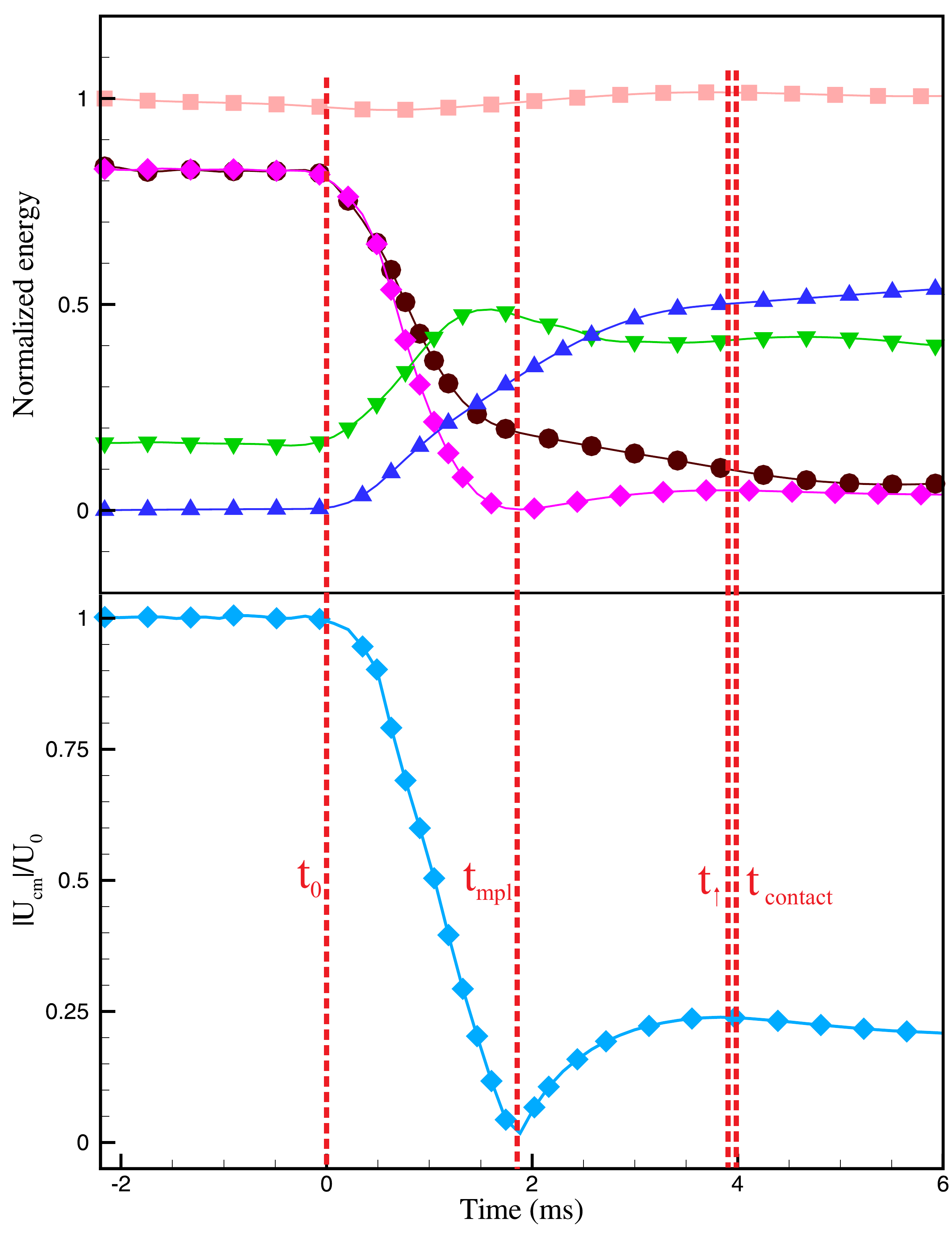}
		\label{figa}}
	\quad
	\subfigure[$We=80$]{ %
		\includegraphics[width=0.4\textwidth] {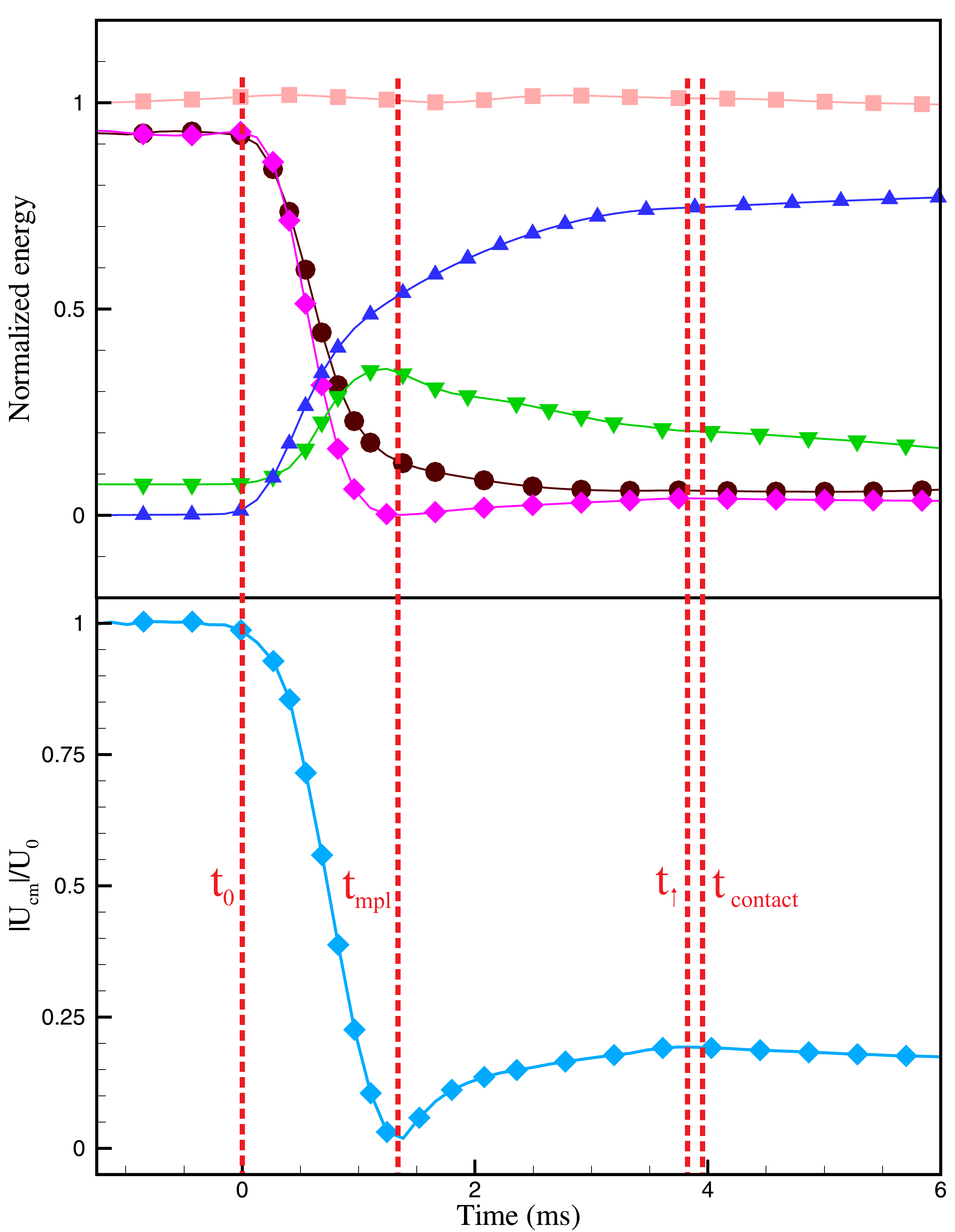}
		\label{figb}}
	\caption{Top panel: History of various components of the energy balance. Circle: Normalized kinetic energy $\tilde{K}$; Downward triangle: Normalized surface energy $\tilde{S}$; Upward triangle: Normalized dissipated energy $\tilde{\Xi}$. Squares: Normalized energy balance $\tilde{K}+\tilde{S}+\tilde{\Xi}$. Diamond:  Normalized center-of-mass kinetic energy $\tilde{K}_{\rm cm}$. 
		Bottom panel: Reduced center-of-mass velocity of the drop $U_{cm}/U_0$. Impact on a perfectly coated SHT $\theta=165^\circ$ for low and high Weber numbers (Supplementary Movie 8). }\label{fig:EnergyWe30_80}
\end{figure}
All the three components of the energy balance equation (\ref{eq:energyconserv}) were evaluated individually for the impact on a perfectly coated texture with the texture density $\Lambda_{\rm exp}=7.25$ (sec.\ \ref{sec:comparison} and \ref{PerfectSur}). 
In Fig.\ \ref{fig:EnergyWe30_80} we present the time evolution of $\tilde{K}$, $\tilde{S}$ and $\tilde{\Xi}$. We also show the normalized center-of-mass kinetic energy $\tilde{K}_{\rm cm}={K}_{\rm cm}/E_0$ where  $K_{\rm cm}=\frac{1}{2}mU_{\rm cm}^2$ is the kinetic energy of center-of-mass, with $m$ the mass of the liquid and $U_{\rm cm}$ the center-of-mass velocity. Results for three representative Weber numbers are shown: ${\rm We}=15$ (shortly after the onset of pancake bouncing at ${\rm We}^*\approx 12$); ${\rm We}=30$ (at the limit of the experimentally accessed Weber numbers; significant intrusion of liquid into the texture) and ${\rm We}=80$ (large intrusion).

Before discussing the results, a brief comment on the validation of the numerics is in order: First, the numerical result at $t=0$ (spherical unperturbed drop moving with the velocity $U_0$) satisfies well the exact relations, $\tilde{K}_0={\rm We}/({\rm We}+6)$, $\tilde{S}_0=6/({\rm We}+6)$: see table \ref{tab:sph}. Also, the energy balance (\ref{eq:energyconserv}) is satisfied within $2\%$ for all times and Weber numbers which is consistent within the accuracy of evaluation of velocity gradients in the computation of energy dissipation.


\begin{table}
\centering
\caption{Reduced kinetic and surface energy for the spherical drop}\label{tab:sph}
\begin{tabular}{l|ll|ll}
	${\rm We}$ & $\tilde{K}_0^{\rm exact}$ & $\tilde{K}_0^{\rm num}$ &  $\tilde{S}_0^{\rm exact}$ &  $\tilde{S}_0^{\rm num}$ \\
	\hline
	$15$       & $0.714$                   & $0.712$                 &  $0.286$                   &  $0.287$\\
	$30$       & $0.833$                   & $0.835$                 &  $0.167$                   &  $0.164$ \\
	$80$       & $0.930$                   & $0.926$                 &  $0.07$                    &  $0.075$
\end{tabular}
\end{table}



The first observation concerns the kinetic energy $\tilde{K}$ and the center-of-mass kinetic energy $\tilde{K}_{\rm cm}$. 
While the latter vanishes at the maximal penetration of the drop into the texture, the kinetic energy itself is different from $\tilde{K}_{\rm cm}$. This difference is attributed to the flow inside the rim of the upper part of the drop remaining above the pillars. The non-negligible amount of kinetic energy carried by the flow of this type was indicated in Ref.\ \cite{clanet2004maximal} for the drop impact on a flat surface. With the increase of Weber number, a greater part of the droplet penetrates the texture, hence the amount of energy in the vortical flow decreases. This is consistent with the result of simulation which shows that the relative difference between $\tilde{K}$ and $\tilde{K}_{\rm cm}$ decreases with the Weber number.

Second, the surface energy $\tilde{S}$ rapidly increases after the impact, as expected.
For ${\rm We}=15$ we see two maxima of $\tilde{S}$, a local maximum close to the zero of the center-of-mass velocity (the drop has stopped penetrating into the texture) and then the global maximum at the time of maximum lateral extension. Note that in this case, maximal extension comes after the droplet bounces off the texture. However, with the increase of the Weber number, the second maximum tends to disappear, and is not present at ${\rm We}=80$. This situation can be termed a pseudo-pancake rebound in order to distinguish it from the “true” one at ${\rm We}=15$: The maximal stretching synchronizes with the maximum penetration time and does not affect the contact time, as was shown above.

Finally, it is clear from the energy balance at all Weber numbers that the dissipation is not negligible in any of the cases for the simulated Ohnesorge number ${\rm Oh}=0.025$. 
While for the the lower ${\rm We=15}$, the surface energy becomes dominant soon after the impact, dissipation is not small even in that case, and levels around $25\%$ at the rebound.  
It is seen from Fig.\ref{fig:EnergyWe30_80} the fraction of energy loss at higher Weber numbers is even larger and dominates others at ${\rm We}=80$. 

Ref.\ \cite{liu2015controlling} discussed a possibility that marcoscopic air pockets get trapped between the droplet and the substrate. Such trapped pockets of air or vapor can undergo compression which could serve as an additional storage of energy to be released into the kinetic energy during the capillary emptying. 
However, the energy balance evaluation above suggests that such a scenario need not be present. 

Summarizing, for the macroscopically flat superhydrophobic surface, the scaling of the contact time holds whenever the Ohnesorge number is not too large (${\rm Oh}<1$) \cite{Antonini2016}. Similar universality holds also in the case of tapered macro-texture: while the dissipation is not necessarily negligible for ${\rm Oh}=0.025$, the contact time of the pancake bouncing scales the same way as in the experiment with water droplets. The only important requirement for that mechanism to be realized is the clear dominance of the surface energy over the kinetic energy at the instance of maximal penetration into the texture. Thus, such energy balance analysis could be very handy in estimating the role played by kinetic, surface and viscous forces for a droplet wall interaction. Such analysis is of greater use when the underlying mechanisms of droplet bouncing are not well understood for example complex macro-textured surfaces. Moreover, imbalance in energy analysis could lead to better understanding of the role played by external factors such as trapped air pockets, thus enabling the design and optimization of novel surface textures.

\subsection{Impact on a tilted texture}

\begin{figure}
\centering
	\includegraphics[width=0.95\textwidth]{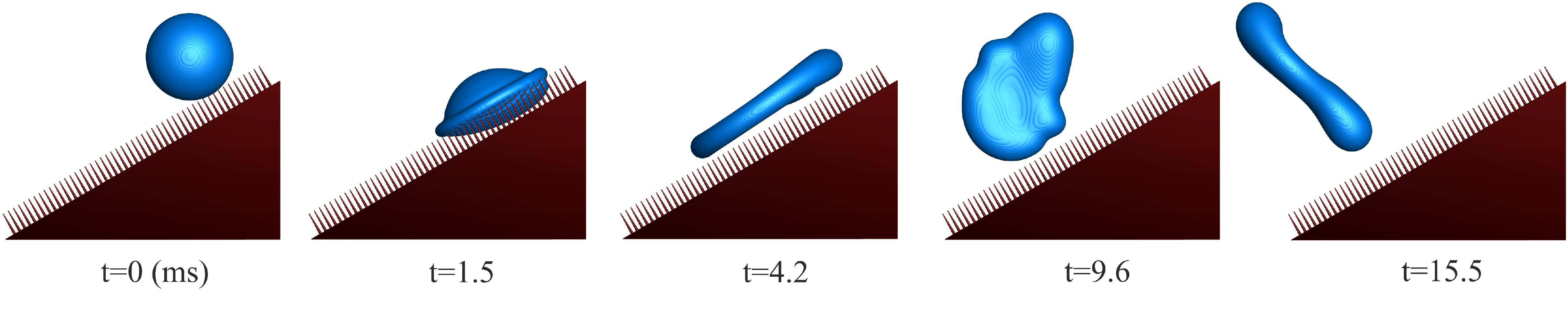}
	\caption{Snapshots of an impact on tapered posts tilted at $30^\circ$;  ${\rm We}=31.2$. The drop rebounds at $t_{\rm contact}=3.6\ [ms]$ which is in excellent agreement with the experiment \cite{liu2014pancake}. 
	Snapshots correspond to Figure 1d of Ref.\ \cite{liu2014pancake} (Supplementary Movie 9). }
	\label{fig:tilt_We31}
\end{figure}

\begin{figure}
\centering
	\includegraphics[width=0.65\textwidth]{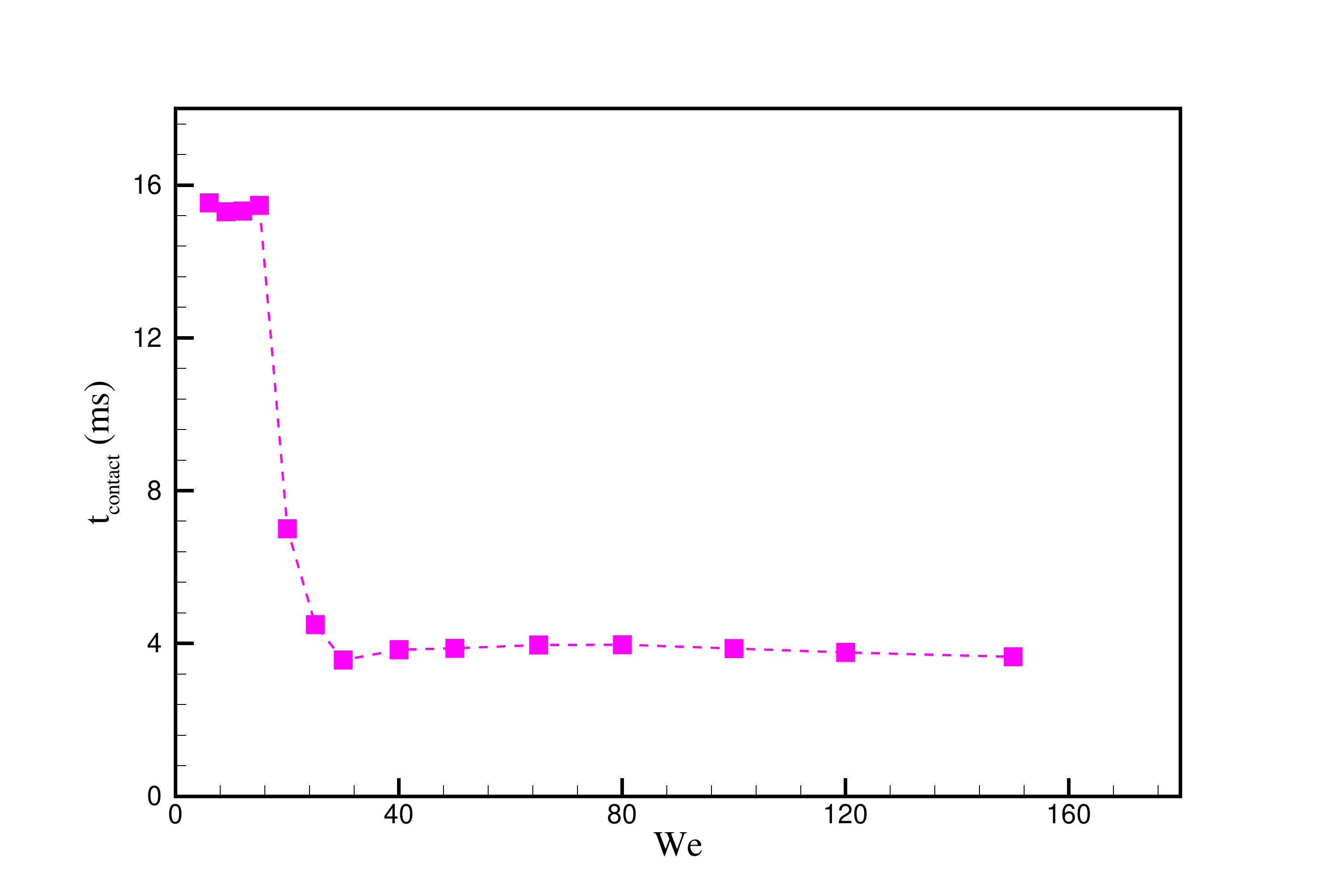}
	\caption{Contact time of a liquid drop impinging on tapered surface under a tilt angle of $\theta=30^o$ for a large range of $We$ ($6\leq We\leq 150$).  Our simulations show that significant reduction in contact time occurs for $We \geq 25$.}
	\label{fig:tcontact_tilt}
\end{figure}

In this section, the simulation results for a drop impacting a tapered texture tilted at $30^\circ$ are presented. 
Fig.\ \ref{fig:tilt_We31} demonstrates that pancake bouncing takes place also for a tilted surface. The shape as well as the contact time ($t_{contact}=3.6\ [ms]$) for the droplet shown in Fig.\ref{fig:tilt_We31} are in good agreement with those observed in the experiment \cite{liu2014pancake}.
Fig.\ \ref{fig:tcontact_tilt} reports the contact time for the tilted surface in a range of Weber numbers from ${\rm We}=6$  to ${\rm We}=150$. 
The pancake bouncing sets in at $We \geq 25$, that is, almost at a twice higher Weber number as compared to the normal impact.
This happens since the motion along the slope delays the penetration of the liquid into the texture. It is also noted that the transition to the pancake bouncing is more gradual than a sharp transition observed for the horizontally aligned substrate.

\section{Conclusions}
\label{sec:cnclude}
The dynamic behavior of a liquid drop impacting a surface with tapered posts was numerically investigated over a wide range of Weber numbers using two-phase entropic lattice Boltzmann method. Superior stability of the ELBM and flexibility of wall boundary conditions allow us to study, for the first time, the pancake bouncing phenomenon in complete detail. Quantitative comparisons of ELBM simulations with previous experiments demonstrate the predictive nature of the multiphase entropic lattice Boltzmann model \cite{PhysRevLett.114.174502}.  

Apart from varying the surface parameters such as the spacing between posts and contact angle, this simulation technique allows us to accurately account for the transformation of kinetic energy into surface energy and vice-versa. We presented numerical evidence that reduction in contact time occurs entirely due to increase of droplet surface area which acts as a storage of kinetic energy during the impact process. Such energy balance analysis, for the first time, allows us to accurately design and optimize surfaces and understand the role played by various physical phenomenon involved in droplet wall interactions. Furthermore the impact of surface superhydrophobic coating can be quantitatively accessed through numerical simulations.

This work was supported by the European Research Council (ERC) Advanced Grant 291094-ELBM and the ETH Research Grant ETH35-12-2.
Computational resources at the Swiss National Super Computing Center CSCS were provided under the grant s492 and s630.

\bibliographystyle{plain}
\bibliography{Ref}

\end{document}